\documentclass[prd, 11pt,a4paper,twocolumn,groupedaddress,preprintnumbers,nofootinbib,floatfix]{revtex4-1}
\usepackage[top=3cm, bottom=2.1cm, left=2cm, right=2cm]{geometry}
\pdfoutput=1

\usepackage{comment}

\usepackage{graphicx}
\usepackage{url}
\usepackage[bookmarks, pagebackref=false]{hyperref}
\usepackage[usenames,dvipsnames]{xcolor}
\usepackage{dcolumn}% Align table columns on decimal point
\usepackage{bm}% bold math
\usepackage{bbm}
\usepackage{amsmath,amssymb,amsfonts}
\usepackage{color}
\usepackage{hyperref}
\usepackage[Symbolsmallscale]{upgreek}
\usepackage{dsfont}
\usepackage[vcentermath]{youngtab}
\usepackage[all]{xy}
\usepackage{pstricks}
\usepackage{dsfont}%
\usepackage{mathtools}
\usepackage{placeins}
\usepackage{enumerate}
\usepackage{cases}
\usepackage{placeins}
\usepackage{xspace}
\usepackage{cancel} % to make the barred text notation
\usepackage{slashed}
\usepackage[caption=false, labelformat=simple, listofformat=subsimple, labelfont=default, margin=5pt,
    justification=raggedright]{subfig}
\usepackage{natbib}
\usepackage{ulem}

\newcommand{\beq}{\begin{eqnarray}}
\newcommand{\eeq}{\end{eqnarray}}

\newcommand{\bmp}{\noindent\begin{minipage}{16cm}}
\newcommand{\emp}{\end{minipage}\vskip 7mm} % 7mm untightened

    \newcommand{\ii}{\mathrm{i}}

    \newcommand{\SU}{\mathrm{SU}} 
    \newcommand{\Sp}{\mathrm{Sp}}
    
    \newcommand{\Tr}{\mathrm{Tr}}

    \newcommand{\SUL}{\mathrm{SU}(2)_{\mathrm{L}}} 
    \newcommand{\SUR}{\mathrm{SU}(2)_{\mathrm{R}}} 
        \newcommand{\vw}{v_{\mathrm{w}}}

    \newcommand{\bee}{\begin{equation}}
        \newcommand{\eee}{\end{equation}}

\def\lsim{\mathrel{\rlap{\lower4pt\hbox{\hskip1pt$\sim$}}
    \raise1pt\hbox{$<$}}}                % less than or approx. symbol
\def\gsim{\mathrel{\rlap{\lower4pt\hbox{\hskip1pt$\sim$}}
    \raise1pt\hbox{$>$}}}                % greater than or approx. symbol

\begin{document}

\title{A partially composite Goldstone Higgs}
\author{Tommi Alanne}
\email{tommi.alanne@mpi-hd.mpg.de}
 \affiliation{Max-Planck-Institut f\"{u}r Kernphysik,\\ Saupfercheckweg 1, 69117 Heidelberg, Germany}
\author{Diogo Buarque Franzosi}
\email{dbuarqu@gwdg.de}
 \affiliation{
 II. Physikalisches Institut, Universit\"at G\"ottingen, Friedrich-Hund-Platz 1, 37077 G\"ottingen, Germany
}
\author{Mads T. Frandsen}
\email{frandsen@cp3.sdu.dk}
 \affiliation{CP$^{3}$-Origins, University of Southern Denmark, Campusvej 55, DK-5230 Odense M, Denmark}

\begin{abstract}
We consider a model of dynamical electroweak symmetry breaking with a partially composite Goldstone Higgs. 
The model is based on a strongly-interacting fermionic sector coupled to a fundamental scalar sector via Yukawa interactions. 
The $\SU(4)\times \SU(4)$ global symmetry of these two sectors is broken to a single $\SU(4)$ via Yukawa interactions. 
Electroweak symmetry breaking is dynamically induced by condensation due to the strong interactions in the new fermionic sector which 
further breaks the global symmetry $\SU(4) \to \Sp(4)$. 
The Higgs boson arises as a partially composite state which is an exact Goldstone boson in the limit where SM interactions are turned off. 
Terms breaking the $\SU(4)$ global symmetry explicitly generate a mass for the Goldstone Higgs. 
The model realizes in different limits both (partially) composite Higgs and (bosonic) Technicolor models, thereby providing a convenient unified framework for phenomenological studies of 
composite dynamics. It is also a dynamical
extension of the recent elementary Goldstone-Higgs model.

{\footnotesize  \it Preprint: 
CP$^3$-Origins-2017-037 DNRF90}
\end{abstract}
\maketitle
\newpage

%%%%%%%%%%%%%%%%%%%%%%%%%%%%%%%%%%%%%%%%%%%%%%%%%%%%%%%%%%%%%%%%%%%%%%%%%%%
\section{Introduction}
\label{sec:intro}
%%%%%%%%%%%%%%%%%%%%%%%%%%%%%%%%%%%%%%%%%%%%%%%%%%%%%%%%%%%%%%%%%%%%%%%%%%%
Gauge-Yukawa models with a strongly interacting fermion sector were proposed in~\cite{'tHooft:1979bh} for 
electroweak symmetry breaking (EWSB) and Standard Model (SM) fermion mass generation. The motivation was to achieve dynamical EWSB
and to alleviate the SM naturalness problem, while circumventing the challenges in building a dynamical 
gauge theory of fermion masses~\cite{Eichten:1979ah,Dimopoulos:1979es,Kaplan:1991dc}. 
Later developments 
of this idea include (partially-)Composite-Higgs
models (pCH) \cite{Kaplan:1983fs,Galloway:2016fuo,Agugliaro:2016clv} and bosonic Technicolor 
(bTC)~\cite{Simmons:1988fu,Samuel:1990dq,Kagan:1990az,Carone:2012cd}. 
In the former case, the SM-Higgs-like scalar can arise as a mixture between a doublet of Goldstone bosons (GBs)
from the composite 
dynamics and an elementary scalar doublet. In the latter case, a SM-Higgs-like scalar can arise as a mixture of an isosinglet composite 
resonance and an elementary scalar. 

Here we study a model with four Weyl fermions transforming under a new $\SU(2)_{\mathrm{TC}}$ gauge group and coupled via 
Yukawa interactions to an $\SU(4)$ multiplet of scalars in the two-index antisymmetric representation. 
This leads to an $\SU(4)/\Sp(4)$ coset structure and a parameter space encompassing both the (p)CH and (b)TC models
while being a dynamical extension of the elementary Goldstone-Higgs model~\cite{Alanne:2014kea}. 

The SM fermions obtain their masses via ordinary Yukawa couplings to the elementary scalar multiplet. 
The weak gauge bosons, on the other hand, obtain masses from both a vacuum expectation value (vev) of the elementary Higgs multiplet and 
the composite sector such that the electroweak (EW) scale $ \vw=246$~GeV is set by
\begin{equation}
    \label{eq:vevs}
    \vw^2=v^2+f^2\sin^2\theta,  
\end{equation}
where $v$ is the vev of the neutral CP-even component of the elementary weak doublet in the $\SU(4)$ multiplet,
$f$ is the GB decay constant of the composite sector, and $\theta$ is the vacuum-misalignment 
angle ($ \pi/2\leq \theta \leq \pi$). 
The TC limit $\theta= \pi/2$,  $v=0$ was studied in e.g.~\cite{Ryttov:2008xe,Luty:2008vs}
while the CH 
limit  $ \pi/2 <\theta < \pi$, $v=0$ was considered in~\cite{Katz:2005au,Gripaios:2009pe,Galloway:2010bp,Barnard:2013zea,Ferretti:2013kya,Cacciapaglia:2014uja,
Franzosi:2016aoo}. Finally the pCH limit, $\pi/2 <\theta < \pi$, $v\neq0$, has recently been studied in~\cite{Galloway:2016fuo,Agugliaro:2016clv}. 
In all cases, the effective low-energy description is based on the 
$\SU(4)/\Sp(4)$ coset.

Since the model here is formulated explicitly 
in terms of elementary constituents, the composite contributions to the spectrum may be predicted using lattice simulations~\cite{Arthur:2016dir,Arthur:2016ozw}. 
Differently from \cite{Kaplan:1983fs,Galloway:2016fuo,Agugliaro:2016clv}, the partially composite Higgs remains an exact GB
in the presence of the Yukawa interactions between the new fermions and the scalar multiplet. 
We find a quartic self-coupling of the scalar multiplet which can be larger 
than in the SM and help alleviate the potential vacuum stability bounds. 
The quartic coupling was set to zero in \cite{Galloway:2016fuo}, and it typically comes out small in bTC models  leading to stringent vacuum stability bounds \cite{Carone:2012cd}. 
Finally, we also discuss the classically scale-invariant limit of the model. 

The paper is organised as follows: In Sec.~\ref{sec:elemComp} we introduce the $\SU(4)$-symmetric model, and in Sec.~\ref{sec:breaking}
we discuss the minimal set of explicit breaking sources leading to the desired phenomenology. In Sec.~\ref{sec:SI} we consider the 
classically scale-invariant limit, and in Sec.~\ref{sec:conclusions} we conclude.

%%%%%%%%%%%%%%%%%%%%%%%%%%%%%%%%%%%%%%%%%%%%%%%%%%%%%%%%%%%%%%%%%%%%%%%%%%%
\section{The Model and the effective description}
\label{sec:elemComp}
%%%%%%%%%%%%%%%%%%%%%%%%%%%%%%%%%%%%%%%%%%%%%%%%%%%%%%%%%%%%%%%%%%%%%%%%%%%

We consider an underlying $\SU(2)_{\mathrm{TC}}$ TC model with four Weyl fermions in the fundamental representation of 
$\SU(2)_{\mathrm{TC}}$. 
The fermion content
in terms of left-handed Weyl fields, with $ \widetilde{U}_{\mathrm{L}}\equiv \epsilon U_R^* $, along with their 
EW quantum numbers is presented in Table~\ref{tab:su4}.
\begin{table}[h!]
    \caption{The new fermion content.}
    \label{tab:su4}
    \begin{center}
	\begin{tabular}{cccc}
	    \toprule
	    & $\vphantom{\frac{\frac12}{\frac12}}\quad\SU(2)_{\mathrm{TC}}\quad$ & $\SU(2)_{\mathrm{W}}\quad$ 
	    & $\mathrm{U}(1)_Y\quad$\\
	    \colrule
	    $\vphantom{\frac{\frac12}{\frac12}} (U_L,D_L)$	&   ${\tiny \yng(1)}$	&   ${\tiny \yng(1)}$	&   0\\    
	    $\vphantom{\frac{1}{\frac12}} \widetilde{U}_{\mathrm{L}}$	&   ${\tiny \yng(1)}$	&   1	&   $-1/2$\\    
	    $\vphantom{\frac{1}{\frac12}} \widetilde{D}_{\mathrm{L}}$	&   ${\tiny \yng(1)}$	&   1	&   $+1/2$ \\
	    \botrule
	\end{tabular}
    \end{center}
\end{table}

With EW and Yukawa interactions turned off, the new fermion sector features a $\SU(4)$ global symmetry, with 
$Q_L=(U_{\mathrm{L}},\, D_{\mathrm{L}},\, \widetilde{U}_{\mathrm{L}},\, \widetilde{D}_{\mathrm{L}}\,)$
transforming in the fundamental representation of this $\SU(4)$. In addition, we introduce an (elementary) scalar  multiplet,
$\Phi$, in the
two-index antisymmetric representation of $\SU(4)$ transforming as $\Phi \to g \Phi g^T$ under $\SU(4)$ transformations. 

The Lagrangian describing the TC and scalar sectors is given by
\begin{equation}
    \label{eq:UVLag}
    \begin{split}
	\mathcal{L}_{\mathrm{PCH}}=&\bar{Q}\ii\slashed{D}Q\\
	+&D_{\mu}\Phi^{\dagger}D^{\mu}\Phi -m_{\Phi}^2\Phi^{\dagger}\Phi-\lambda_\Phi(\Phi^{\dagger}\Phi)^2\\
	-&y_Q   Q^T \Phi Q+\mathrm{h.c.},
    \end{split}
\end{equation}
where the antisymmetric flavour structure of the fermion bilinear, after antisymmetrising Lorentz and gauge indices, is kept implicit.

This Lagrangian is $\SU(4)$ invariant in the limit $g,g' \to 0$ and further $\SU(4)_Q\times \SU(4)_\Phi$ invariant when in addition $y_Q=0$ 
with the two copies of $\SU(4)$ acting on the fermionic and scalar sectors, respectively. 
Furthermore, it is scale invariant at the classical level if $m_{\Phi}^2=0$.

We first consider the fermionic sector in the limit $y_Q=0$.  
The flavour-antisymmetric condensate 
$\langle\epsilon Q_I^T Q_J \rangle \sim  f^3 E_{IJ} $ breaks $\SU(4)_Q$ to $\Sp(4)_Q$. We embed $\SUL\times\SUR$ in 
$\SU(4)_Q$ by identifying the left and right generators
\begin{equation}
    \label{eq:gensLR}
    T^i_{\mathrm{L}}=\frac{1}{2}\left(\begin{array}{cc}\sigma_i & 0 \\ 0 & 0\end{array}\right),\
    T^i_{\mathrm{R}}=\frac{1}{2}\left(\begin{array}{cc} 0 & 0 \\ 0 & -\sigma_i^{T}\end{array}\right),
\end{equation}
where $\sigma_i$ are the Pauli matrices.  The EW subgroup is gauged after identifying the generator of hypercharge with 
$T_{\mathrm{R}}^3$; see e.g.~\cite{Luty:2004ye,Cacciapaglia:2014uja,Franzosi:2016aoo} for details. 

The breaking of the EW gauge group depends on the alignment between the EW subgroup and the stability group $\Sp(4)_Q$.
This can be conveniently parameterized by an angle, $\theta$, after identifying the vacua that leave the 
EW symmetry intact, $E_\pm$, and the one, $E_{\mathrm{B}}$, breaking it completely to $\mathrm{U}(1)_{\mathrm{EM}}$ 
of electromagnetism. 
The true vacuum can be written as a linear combination of the EW preserving and breaking vacua, 
$E=\cos\theta E_-+\sin\theta E_{\mathrm{B}}$. Either choice of $E_{\pm}$ is equivalent, and here we have chosen 
$E_-$. These different vacua can be written explicitly \cite{Galloway:2010bp} as
\begin{equation}
    E_\pm = \left( \begin{array}{cc}
	\ii \sigma_2 & 0 \\
	0 & \pm \ii \sigma_2
    \end{array} \right)\,,\qquad
    E_B  =\left( \begin{array}{cc}
	0 & 1 \\
	-1 & 0
    \end{array} \right) \, ,
\end{equation}
and the angle $\theta$ is that in Eq.~(\ref{eq:vevs}).

The composite Goldstone degrees of freedom in the $\SU(4)/\Sp(4)$ coset can be parameterised by the 
exponential map
\begin{equation}
    \label{eq:composite}
    \Sigma=\exp\left(\frac{2\sqrt{2}\,\ii}{f}\Pi^a X^a\right)E,
\end{equation}
where the $X^a$, with $a=1,\dots,5$, are the broken generators corresponding to the vacuum $E$.

We now include the scalar sector and parameterise the scalar multiplet in terms of EW eigenstates according to the vacuum $E_-$. 
To incorporate the EW embedding, we introduce spurions, $P_{\alpha}$, $\tilde{P}_{\alpha}$, 
\begin{equation}
    \label{eq:proj}
    \begin{split}
	2P_{1}&=\delta_{i1}\delta_{j3}-\delta_{i3}\delta_{j1},\ 
	2P_{2}=\delta_{i2}\delta_{j3}-\delta_{i3}\delta_{j2},\\
	2\tilde{P}_{1}&=\delta_{i1}\delta_{j4}-\delta_{i4}\delta_{j1},\ 
	2\tilde{P}_{2}=\delta_{i2}\delta_{j4}-\delta_{i4}\delta_{j2},
    \end{split}
\end{equation}
such that $H_\alpha \equiv \Tr[P_{\alpha}\Phi]$ and $\tilde{H}_\alpha\equiv \Tr[\tilde{P}_{\alpha}\Phi]$ transfrom as EW doublets with 
hypercharges $+1/2$ and $-1/2$, respectively. 
Furthermore, the projectors to the EW-singlet directions are given by
\begin{equation}
    \label{eq:}
    P^{\mathrm{S}}_1=\frac{1}{2}\left(\begin{array}{cc}-\ii\sigma_2 & 0\\ 0 & 0\end{array}\right),\ 
    P^{\mathrm{S}}_2=\frac{1}{2}\left(\begin{array}{cc}0 & 0\\ 0 & \ii\sigma_2\end{array}\right).
\end{equation}
In terms of the above projectors, we can write the scalar multiplet as
\begin{equation}
    \label{eq:Phidef}
    \Phi=\sum_{\alpha=1,2}P_{\alpha}H_\alpha+\tilde{P}_{\alpha}\widetilde{H}_{\alpha}
	+P^{\mathrm{S}}_1 S+P^{\mathrm{S}}_2 S^{*},
\end{equation}
where 
\begin{equation}
    \label{eq:H}
    H=\frac{1}{\sqrt{2}}\left(\begin{array}{c}\sigma_h-\ii \pi_h^3 \\ -(\pi_h^2+\ii \pi_h^1)\end{array}\right),
\end{equation}
and $S=\frac{1}{\sqrt{2}}(S_{\mathrm{R}}+\ii S_{\mathrm{I}})$ parameterises the EW-singlet scalars. 

We couple the SM fermions, in particular the top quark, to the EW-doublet scalar $H$ via the standard Yukawa interactions of the form 
$y_t \bar{q}_{\mathrm{L}} H t_{\mathrm{R}}+\text{ h.c.}$ where $q_L=(t_L,b_L)$.

Turning on $y_Q$ breaks the 
$\SU(4)_Q\times \SU(4)_\Phi\to \SU(4)$, and upon condensation of the technifermions the $\SU(4)$ spontaneously breaks to $\Sp(4)$. 
The five GBs may be classified as an EW-doublet Goldstone Higgs and an EW singlet. 

Below the condensation scale, the Lagrangian of Eq.~\eqref{eq:UVLag} yields
\begin{equation}
    \label{eq:effLag}
    \mathcal{L}_\mathrm{eff}=\mathcal{L}_{\mathrm{kin}}-V_{\mathrm{eff}},
\end{equation}
where after gauging the EW subgroup, the kinetic terms read
\begin{equation}
    \label{eq:kinLag}
    \mathcal{L}_{\mathrm{kin}}=\Tr [D_{\mu}\Phi^{\dagger}D^{\mu}\Phi]+\frac{f^2}{8}\Tr[D_{\mu}\Sigma^{\dagger}D^{\mu}\Sigma],
\end{equation}
with 
\begin{equation}
    \label{eq:covD}
    D_{\mu}X=\partial_{\mu}X-\ii\left(G_{\mu}X+XG_{\mu}^{T}\right),
\end{equation}
$X=\Phi,\Sigma$, and the EW gauge fields are encoded in
\begin{equation}
    \label{eq:Gmu}
    G_{\mu}=g W_{\mu}^iT_{\mathrm{L}}^i+g^{\prime}B_{\mu}T_{\mathrm{R}}^3.
\end{equation}
Furthermore, the effective tree-level potential  is given by
\begin{equation}
    \label{eq:Veff}
    \begin{split}
	V_{\mathrm{eff}}=&4\pi f^3Z_2\left(y_Q \Tr\left[\Phi\Sigma\right]+\ \text{h.c.}\right)\\
	&+m_{\Phi}^2\Tr[\Phi^{\dagger}\Phi]+\lambda_{\Phi}\Tr[\Phi^{\dagger}\Phi]^2,
    \end{split}
\end{equation}
where $Z_2$ is a non-perturbative constant expected to be $O(1)$. We use the numerical value 
$Z_2\approx 1.5$ following from the determination of the condensate, in the $\SU(2)$ gauge theory with two flavours,  in~\cite{Arthur:2016dir}.

Then the vacuum conditions read
\begin{equation}
    \label{eq:SU4ComplVac}
    y_Q=\frac{m^2_{\lambda} v}{8\sqrt{2}\pi Z_2 f^3 s_{\theta}},\quad t_{\theta}=-\frac{v}{v_S},
\end{equation}
where $v\equiv\langle\sigma_h\rangle$, $v_S\equiv \langle S_{\mathrm{R}}\rangle$, 
and $m^2_{\lambda}\equiv m_{\Phi}^2+\lambda_{\Phi}(v^2+v_S^2)$. We use the shorthand notations $s_x\equiv \sin x, c_x\equiv \cos x, t_x\equiv \tan x$
throughout the paper.

The CP-even neutral states $\sigma_h, \Pi_4, S_{\mathrm{R}}$ mix.  Using Eq.~\eqref{eq:vevs} and
\begin{equation}
    \label{eq:tbeta}
    t_{\beta}\equiv \frac{v}{f s_{\theta}},
\end{equation}
we can write the mass eigenvalues as
\begin{equation}
    \label{eq:masses0}
    \begin{split}
	m_1^2&=0,\ m_2^2=\frac{m_{\lambda}^2}{c_{\beta}^2},\ m_3^2=m^2_{\lambda}+2\lambda_{\Phi}\vw^2\frac{s_{\beta}^2}{s_{\theta}^2}.
    \end{split}
\end{equation}
The eigenstates corresponding to the eigenvalues are given by
\begin{equation}
    \label{eq:eigVect0}
    \begin{split}
	h_1=&s_{\beta}c_{\theta}\sigma_h+c_{\beta}\Pi_4+s_{\beta}s_{\theta}S_{\mathrm{R}},\\
	h_2=&c_{\beta}c_{\theta}\sigma_h-s_{\beta}\Pi_4+c_{\beta}s_{\theta}S_{\mathrm{R}},\\
	h_3=&s_{\theta}\sigma_h-c_{\theta}S_{\mathrm{R}} . 
    \end{split}
\end{equation}
In the limit of $t_\beta\gg 1$ and $s_\theta\ll 1$ the massless $h_1$ state is mostly $\sigma_h$, the scalar excitation of the elementary doublet. 
The additional four massless states are the would-be GBs eaten by the $W$ and $Z$ bosons as well as an additional CP-odd state.

%%%%%%%%%%%%%%%%%%%%%%%%%%%%%%%%%%%%%%%%%%%%%%%%%%%%%%%%%%%%%%%%%%%%%%%%%%%
\section{Breaking of the global SU(4) and a pseudo-Goldstone Higgs}
\label{sec:breaking}
%%%%%%%%%%%%%%%%%%%%%%%%%%%%%%%%%%%%%%%%%%%%%%%%%%%%%%%%%%%%%%%%%%%%%%%%%%%

The gauging of the EW subgroup and the Yukawa interactions between $H$ and the SM fermions break the global symmetry explicitly. The dominant EW effect comes from 
gauge-boson loops with the techniquarks. The one-loop corrections to the elementary scalar potential from the 
EW and SM-fermion sectors, i.e. the Coleman--Weinberg potential \cite{Coleman:1973jx},
are higher order in perturbation theory.
We include the leading EW contribution in the effective potential
by adding the effective operators~\cite{Peskin:1980gc,Preskill:1980mz}
\begin{equation}
    \label{eq:EWgauge}
    \begin{split}
    V_{\mathrm{gauge}}=&-C_g\left [ g^2f^4\sum_{i=1}^3 \mathrm{Tr}\left(T_{\mathrm{L}}^i\Sigma 
	(T_{\mathrm{L}}^i\Sigma)^*\right)\right.\\
    &\qquad\quad\left.\vphantom{\sum_{i=1}^3}+g^{\prime\,2}f^4 \mathrm{Tr}\left(T_{\mathrm{R}}^3\Sigma
	(T_{\mathrm{R}}^3\Sigma)^*\right)\right ],
    \end{split}
\end{equation}
where $C_g$ is a positive loop-factor, and we expect $C_g\sim\mathcal{O}(1)$.
For the effective potential, these yield 
    \begin{equation}
	\label{eq:}
	V_{\mathrm{eff}}\supset-\frac{1}{2}\widetilde{C}_g Z_2^2f^4c_{\theta}^2,
    \end{equation}
    where $\widetilde{C}_g\equiv C_g(3g^2+g^{\prime\,2})$.

Another possible source of explicit breaking of the global symmetry is splitting of the masses of the EW-doublet and singlet 
components of the scalar multiplet, $\Phi$. We expect a small splitting from quantum corrections from the top-quark and 
EW gauge sectors but here we simply add an explicit mass resulting in such a splitting,
\begin{equation}
    \label{eq:}
    \begin{split}
	V_{\delta m^2}=&2\delta m^2\, \Tr[P_i^S\Phi]\Tr[P_i^S\Phi]^*\\
	=&\frac{1}{2}\delta m^2(S_{\mathrm{R}}^2+S_{\mathrm{I}}^2).
    \end{split}
\end{equation}

We note that the mass of the fifth CP-odd GB state may be lifted by adding two independent mass terms for the singlets without affecting vacuum alignment.

    %%%%%%%%%%%%%%%%%%%%%%%%%%%%%%%%%%%%%%
    \subsection{Vacuum and spectrum}
    %%%%%%%%%%%%%%%%%%%%%%%%%%%%%%%%%%%%%%

    Taking the above sources of explicit $\SU(4)$ breaking into account, the vacuum conditions now read (cf. Eq.~\eqref{eq:SU4ComplVac})
    \begin{equation}
	\label{eq:SU4BrVac}
	\begin{split}
	    y_Q=&\frac{m^2_{\lambda} v}{8\sqrt{2}\pi Z_2 f^3 s_{\theta}}\,,\\
	    v_S=&\frac{\widetilde{C}_g Z_2^2f^4s_{\theta}^2-v^2m_{\lambda}^2}{t_{\theta}vm_{\lambda}^2}\,,\\
	    \delta m^2=&\frac{\widetilde{C}_g Z_2^2f^4s_{\theta}^2m_{\lambda}^2}{v^2m_{\lambda}^2-\widetilde{C}_g Z_2^2f^4s_{\theta}^2}\,.
	\end{split}
    \end{equation}
    The CP-even neutral states $\sigma_h, \Pi_4, S_{\mathrm{R}}$ again mix, {but due to the explicit breaking of $\SU(4)$, the 
    lightest mass eigenstate $h_1$ now acquires a non-zero mass and becomes a pseudo-GB (pGB).
    We fix the mass of this lightest state to the 
    observed Higgs mass, $m_{h_1}=125$~GeV,
    and  show the values of the Lagrangian mass parameters $f,\delta m^2, m_\Phi^2$ for fixed values of $s_{\theta}=0.1,0.3$ and $C_g=1$ in 
    the $(\lambda_{\Phi},t_{\beta})$ plane 
    in Figs.~\ref{fig:massPars} and~\ref{fig:massPars2}, respectively. Below the cyan line in the upper panel, $m_\Phi^2>0$, while $\delta m^2>0$ throughout 
    the parameter space as shown in the lower panel. 
    In particular, we can achieve correct EWSB driven purely by the strong dynamics $m_\Phi^2>0,\delta m^2>0$. Furthermore, 
    we generically have $\delta m^2\sim m_\Phi^2\sim f^2$, and for $s_{\theta} \lesssim 0.1$ it is possible to reach scalar mass parameters above a TeV.

    The ratio $m_{\Phi}^2/f^2$ can be written for $s_{\theta}\ll 1$ as
    \begin{equation}
	\label{eq:moverf}
	\begin{split}
	    \frac{m_{\Phi}^2}{f^2}=&\frac{8\sqrt{2}\pi Z_2 y_Q}{t_{\beta}}\\
	    &-\lambda_{\Phi}\left(t_{\beta}^2+\frac{\tilde{C}_g^2Z_2^2}{128\pi^2 y_Q^2}-\frac{\tilde{C}_gZ_2t_{\beta}}{4\sqrt{2}\pi y_Q}\right).
	\end{split}
    \end{equation}
    If we require $t_{\beta}> 1.3$ to avoid the Landau pole for the top-Yukawa coupling and the flavour-changing 
    neutral currents from the heavy pseudoscalars and restrict $\lambda_{\Phi}<1$, then the Higgs mass constraint $m_{h_1}=125$~GeV 
    results in $y_Q\lesssim 0.1$ and further $m_{\Phi}^2/f^2\lesssim  4$ with the maximum at large $\lambda_{\Phi}$ and small 
    $t_{\beta}$.

    The masses of the heavy pion triplet can be written
    as
    \begin{equation}
	\label{eq:piMass}
    m_{\pi}^2=\frac{8\sqrt{2}\pi Z_2\vw^2 y_Q}{t_{\beta}s_{\theta}^2},   
    \end{equation}
    and they are of the same order as the heavy scalar masses.
    The additional scalars are also relatively heavy scaling roughly with $\delta m^2$. The two remaining CP-odd states in the spectrum are the 
    mass eigenstates composed of $\Pi_5$ and $S_{\mathrm{I}}$.

   \begin{figure}
	\begin{center}
	    \includegraphics[width=0.48\textwidth]{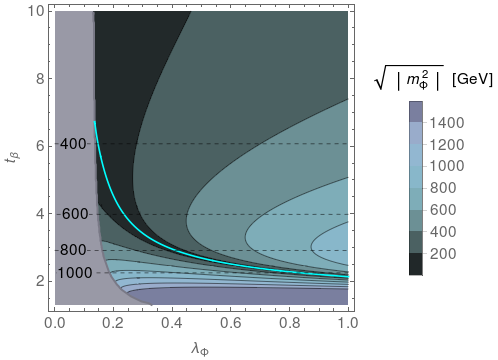}\\
	    \vspace{0.2cm}
	    \includegraphics[width=0.48\textwidth]{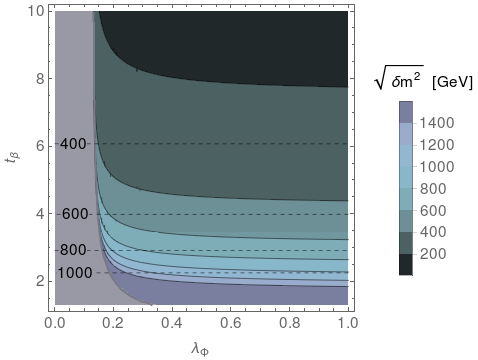}
	\end{center}
	\caption{Contours of the absolute value of the mass parameter $\sqrt{|m_{\Phi}^2|}$ ($\sqrt{\delta m^2}$) is depicted in the upper (lower) panel 
	for a fixed values of $s_{\theta}=0.1$ and $C_g=1$. The mass of the lightest scalar eigenstate is fixed to 125~GeV. On the gray 
	shaded region on the left, no solution for the 
	Higgs mass condition is found. In the upper panel, below the solid cyan curve, $m_{\Phi}^2>0$. In the lower panel, $\delta m^2$ 
	is always positive. Dashed lines indicate the value of the composite-pion decay constant, $f$, 
	in GeV.}
	\label{fig:massPars}
    \end{figure}
  
    \begin{figure}
	\begin{center}
	    \includegraphics[width=0.48\textwidth]{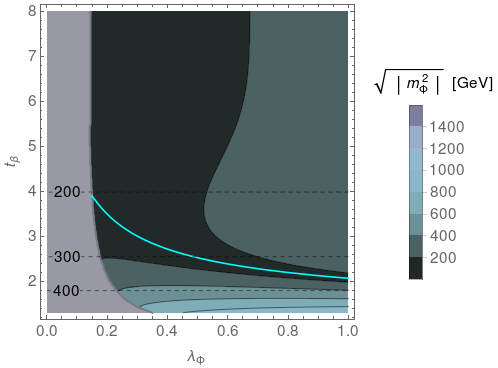}\\
	    \vspace{0.2cm}
	    \includegraphics[width=0.48\textwidth]{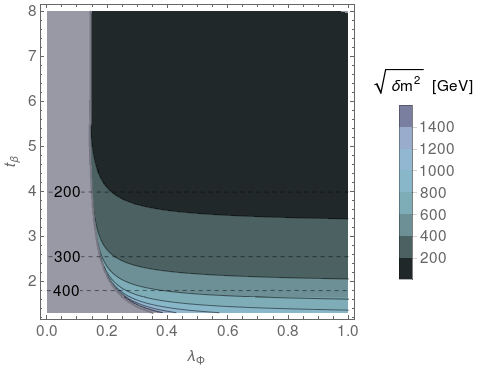}
	\end{center}
	\caption{Contours of the absolute value of the mass parameter $\sqrt{|m_{\Phi}^2|}$ ($\sqrt{\delta m^2}$) is depicted 
	in the upper (lower) 
	panel for a fixed values of $s_{\theta}=0.3$ and $C_g=1$. The mass of the lightest scalar eigenstate $h_1$ is fixed to 125~GeV. 
	On the gray shaded region on the left, no solution for the 
	Higgs mass condition is found. In the upper panel, below the solid cyan curve, $m_{\Phi}^2>0$. In the lower panel, $\delta m^2$ is 
	always positive. Dashed lines indicate the value of the composite-pion decay constant, $f$, in GeV.}
	\label{fig:massPars2}
    \end{figure}

    %%%%%%%%%%%%%%%%%%%%%%%%%%%%%%%%%%%%%%
    \subsection{Couplings}
    %%%%%%%%%%%%%%%%%%%%%%%%%%%%%%%%%%%%%%
    To study the viability of the model in light of the current experimental data, we parameterize the rotation to the mass eigenbasis by
    \begin{equation}
	\label{eq:}
	\left(\begin{array}{c}
	h_1\\
	h_2\\
	h_3
	\end{array}\right)=
	R\left(\begin{array}{c}
	\sigma_h\\
	\Pi_4\\
	S_{\mathrm{R}}
	\end{array}\right),
    \end{equation}
    and define the coefficients
    \begin{equation}
	\label{eq:kappat}
	\kappa_t\equiv\frac{g_{h_1\bar{t}t}}{g_{h\bar{t}t}^{\mathrm{SM}}}=\frac{y_t R_{11}}{y_t^{\mathrm{SM}}}=\frac{R_{11}}{s_{\beta}},
    \end{equation}
    and
    \begin{equation}
	\label{eq:}
	\kappa_V\equiv\frac{g_{h_1W^+W^-}}{g_{hW^+W^-}^{\mathrm{SM}}}=R_{11}s_{\beta}+R_{12}c_{\beta}c_{\theta}.
    \end{equation}
    We solve the $R$ matrix numerically and show the values of $\kappa_t$-coefficient for fixed values of $s_{\theta}=0.1$ and $C_g=1$ in the 
    $(\lambda_{\Phi},t_{\beta})$ plane in Fig.~\ref{fig:kappas}; the values of $\kappa_V$ are almost identical, and therefore we do not show them 
    separately here.
    
    The LHC experiments constrain the couplings of the lightest scalar eigenstate, the Higgs boson, to top quark and EW 
    gauge bosons. The combined ATLAS \& CMS analysis of the Run-1 data~\cite{Khachatryan:2016vau} limits  the modifications of 
    the vector and fermion couplings in a  
    two-parameter fit of $\kappa_f=\kappa_t$ and $\kappa_V$ to less than 20\% at $1\sigma$ level. It is evident that in this respect the model 
    is viable in most of the parameter space. 
    
    \begin{figure}
	\begin{center}
	    \includegraphics[width=0.48\textwidth]{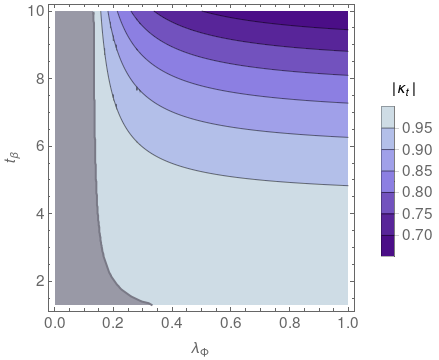}\\
	\end{center}
	\caption{The (absolute) value of the coeffcient $|\kappa_t|$  
	is depicted for a fixed values of $s_{\theta}=0.1$ and $C_g=1$. The mass of the lightest scalar eigenstate is fixed to 125~GeV. 
	On the gray shaded region on the left, no solution for the Higgs mass condition is found.}
	\label{fig:kappas}
    \end{figure}

%%%%%%%%%%%%%%%%%%%%%%%%%%%%%%%%%%%%%%%%%%%%%%%%%%%%%%%%%%%%%%%%%%%%%%%%%%%
  \section{The Classically scale-invariant limit}
    \label{sec:SI}
%%%%%%%%%%%%%%%%%%%%%%%%%%%%%%%%%%%%%%%%%%%%%%%%%%%%%%%%%%%%%%%%%%%%%%%%%%%

Figs.~\ref{fig:massPars} and~\ref{fig:massPars2} show } that it is possible to have $m_\Phi^2=0$, as along the cyan line. In the limit where 
both $m_\Phi^2=\delta m^2=0$, the Lagrangian in Eq.~\eqref{eq:UVLag} then becomes scale invariant at the classical level. Even though the scale invariance 
is broken at the quantum level, the classical scale invariance has been invoked as a guiding principle for the EWSB sector. The model here is distinct from 
models~\cite{Coleman:1973jx,Hempfling:1996ht,Chang:2007ki,Englert:2013gz,Antipin:2013exa} relying on the Coleman--Weinberg 
mechanism~\cite{Coleman:1973jx} to generate EWSB via  one-loop corrections to the elementary scalar potential from the 
EW and SM-fermion sectors. 

Here, EWSB is again induced and communicated to the elementary scalars due to condensation of the technifermions. 
This is similar to the scale-invariant model of~\cite{Heikinheimo:2013fta,Antipin:2014qva} where a singlet scalar is coupled both via Yukawa 
interactions to new strongly-interacting fermions without SM charges and via quartic couplings to the Higgs. 

The  corrections from the EW gauge-boson loops to the techniquarks are again important, and we include these contributions 
to the effective potential as in Eq.~\eqref{eq:EWgauge}.     

To obtain a non-zero mass for the pGB Higgs, we need to add a source of explicit global symmetry
breaking. Contrary to the previous discussion, in the scale-invariant framework we add a splitting between 
the singlet and doublet quartic couplings instead of mass splitting. This is achieved by adding
to the effective potential a term  
\begin{equation}
    \label{eq:Veff}
	V_{\delta \lambda}=\delta \lambda \Tr\left[\Phi_S^{\dagger}\Phi_S\right]^2,
\end{equation}
where $\Phi_S=P^{\mathrm{S}}_1 S+P^{\mathrm{S}}_2 S^{*}$. Due to the different RG running of the EW-singlet and -doublet parts, some splitting
of the quartics is anticipated, and we also note that the RG running would further induce splitting of the form 
$\Tr\left[H^{\dagger}H\right]\Tr\left[\Phi_S^{\dagger}\Phi_S\right]$. Here we remain agnostic about the origin of the splitting and consider the minimal scenario in Eq.~\eqref{eq:Veff}.

\begin{figure}
    \begin{center}
	\includegraphics[width=0.46\textwidth]{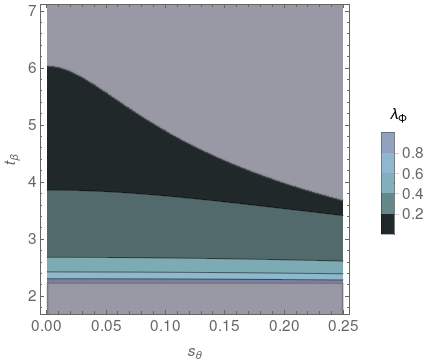}\\
	    \vspace{0.2cm}
	\includegraphics[width=0.46\textwidth]{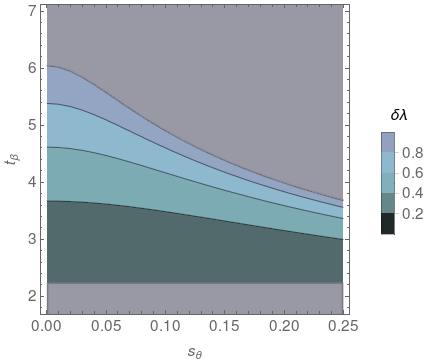}
    \end{center}
    \caption{The values of the $\SU(4)$-invariant quartic coupling, $\lambda_{\Phi}$, (upper panel) and the $\SU(4)$-breaking one, 
    $\delta\lambda$, (lower panel) for fixed value $C_{g}=1$. The mass of the lightest scalar eigenstate, $h_1$, is fixed to 125~GeV. 
    On the gray shaded regions, one of the couplings become large ($\gtrsim 1$).}
    \label{fig:SI2}
\end{figure}

Minimising the potential then leads to 
\begin{equation}
    \label{eq:}
    \begin{split}
	y_Q=&\frac{\widetilde{C}_g Z_2 f s_{2\theta}}{16\sqrt{2}\pi\left(v c_{\theta}+v_S s_{\theta}\right)},\\
	\lambda_{\Phi}=&\frac{\vphantom{\frac{\frac12}{2}}\widetilde{C}_g Z_2^2 f^4 s_{2\theta}s_{\theta}}{2v\left(v^2+v_S^2\right)
	    \left(v c_{\theta}+v_S s_{\theta}\right)},\\
	\delta \lambda=&-\frac{\vphantom{\frac{\frac12}{2}}\widetilde{C}_g Z_2^2 f^4 s_{2\theta}}{2v v_S^3}.
    \end{split}
\end{equation}
Fixing again the mass of the lightest neutral scalar eigenstate to $125$~GeV, we show the values of the scalar quartics, and 
the Yukawa coupling fixing $C_g=1$ in the $(s_{\theta},t_{\beta})$ plane in Fig.~\ref{fig:SI2}. 
We disregard the grey areas where the quartic couplings are larger than one, and we 
again find viable parameter space for a range of $t_\beta$ and $s_\theta$ values. 
For relatively small $t_{\beta}$, 
it is possible to have $\delta\lambda<\lambda_{\Phi}<1$. As in the previous case, the Yukawa couplings must be small at 
the $O(10^{-2}\dots10^{-1})$ level.   

We leave a more detailed study of the phenomenology of the model presented here to future work. 
We also note that additional source of $\SU(4)$ breaking is needed to lift the mass of the remaining CP-odd GB. Minimally
this can be achieved by adding a term of the form $\delta\lambda_S S_R^2S_I^2$. Adding such a term does not affect the 
previous results, and we leave a detailed investigation of the heavy scalar spectrum for future studies. 

%%%%%%%%%%%%%%%%%%%%%%%%%%%%%%%%%%%%%%%%%%%%%%%%%%%%%%%%%%%%%%%%%%%%%%%%%%%
\section{Conclusions}
\label{sec:conclusions}
%%%%%%%%%%%%%%%%%%%%%%%%%%%%%%%%%%%%%%%%%%%%%%%%%%%%%%%%%%%%%%%%%%%%%%%%%%%
In this paper we have proposed a model of a partially composite Goldstone Higgs. 
The model combines dynamical EWSB with SM fermion mass generation via ordinary Yukawa couplings to a scalar multiplet. The model 
features a spontaneously broken $\SU(4)$ global symmetry with the Higgs as a Goldstone boson in the limit of no SM interactions. The Higgs state is mostly 
elementary, and thus its couplings to both the SM-vector and -fermion states are SM Higgs like, while we find a Higgs self-coupling that can be larger than 
in the SM and thereby help alleviate the vacuum stability constraints. 
This is in contrast to the fully elementary realisation, where the quartic coupling is expected to be smaller than in the SM~\cite{Alanne:2014kea}, 
thereby providing a potential diagnostics to distinguish these different realisations in future collider experiments. 

The $\SU(4)$ symmetry is broken by the EW interactions and by adding explicit breaking terms to the scalar potential. We find a 
viable solution with a SM-like Higgs state from an $\SU(4)$-breaking mass term which is small compared to the compositeness scale $4 \pi f$. 
The $\SU(4)$-preserving mass parameter $m_\Phi^2$ remains of the order of $f^2$, but can be positive as opposed to in the SM, and the symmetry breaking can be fully
dynamical. We also find a viable solution in the classically scale-invariant limit 
by setting $m_\Phi^2=0$ and adding $\SU(4)$-non-invariant quartic couplings to the scalar potential.  
In this case the $\SU(4)$-breaking quartic coupling may be smaller than the $\SU(4)$-preserving one.

In all cases, the additional $\SU(4)$ breaking (on top of the SM gauge and Yukawa interactions) can be attained on the scalar-potential level, while the 
new fermion sector remains $\SU(4)$ symmetric.

\section*{Acknowledgments}
We thank M. J{\o}rgensen, M.L.A. Kristensen, C. Pica and K. Tuominen for discussions. 
TA acknowledges partial funding from a Villum foundation grant when part of this article was being completed.
 MTF acknowledges partial funding from The Council For Independent Research, grant number 
DFF 6108-00623. The CP3-Origins centre is partially funded by the Danish National Research Foundation, grant number DNRF90.  
%   

%%%%%%%%%%%%%%%%%%%%%%%%%%%%%%%%%%%%%%%%%%%%%%%%%%%%%%%%%%%%%%%%%%%%%%%%%%%%%%%%%%%%%%%%%%%%%%%%%%%%
%
\bibliography{PCH.bib}
%
%%%%%%%%%%%%%%%%%%%%%%%%%%%%%%%%%%%%%%%%%%%%%%%%%%%%%%%%%%%%%%%%%%%%%%%%%%%%%%%%%%%%%%%%%%%%%%%%%%%%    
    
\end{document}